\documentclass[preprint2]{aastex}
\slugcomment{2017.1 v2}
\usepackage{amsmath}	

\shorttitle{Evolution of pulsar braking index} \shortauthors{Tong \& Kou}

\begin{document}

\title{Possible evolution of the pulsar braking index from larger than three to about one}

\author{H. Tong\altaffilmark{1,2}, F. F. Kou\altaffilmark{2}}
\altaffiltext{1}{School of Physics and Electronic Engineering, Guangzhou University, 510006 Guangzhou, China;
\\ htong\_2005@163.com}
\altaffiltext{2}{Xinjiang Astronomical Observatory, Chinese Academy of Sciences, Urumqi, Xinjiang 830011,
    China}

\begin{abstract}
The coupled evolution of pulsar rotation and inclination angle in the wind braking model is calculated. The oblique pulsar tends to align. The pulsar alignment will affect its spin-down behavior. As a pulsar evolves from the magneto-dipole radiation dominated case to the particle wind dominated case, the braking index will first increase and then decrease. In the early time, the braking index may be larger than $3$. And during the following long time, the braking index will be always smaller than $3$. The minimum braking index is about one. This can explain the existence of high braking index larger than $3$, and low braking index of pulsars simultaneously. The pulsar braking index is expected to evolve from larger than three to about one. A general trend is that the pulsar braking index will evolve from the Crab-like case to the Vela-like case.
\end{abstract}


\section{Introduction}

The pulsar braking index reflects the slow-down law of pulsars (Lyne et al. 2015): $\dot{\nu} \propto -\nu^n$, where $\nu$ and $\dot{\nu}$ are the frequency and frequency derivative, respectively, and $n$ is the so-called braking index. Previously, eight young pulsars have braking index reported (Lyne et al. 2015). Their values are all smaller than three, which lies between $0.9-2.84$. The braking index may be very useful to discriminate between different spin-down mechanisms of pulsars. A braking index of three is predicted by the magnetic dipole braking assumption. The possible explanations for a braking index smaller than three include: the presence of fallback disks (Liu et al. 2014), an increasing magnetic field (Espinoza et al. 2011),  an increasing pulsar inclination angle (Lyne et al. 2013), or particle outflow in the magnetosphere (e.g., wind braking of pulsars, Kou \& Tong 2015) etc. 

There is marginal evidence for the evolution of pulsar braking index (Espinoza 2013). Recently, one pulsar PSR J1640$-$4631 is reported to have a braking index larger than three: $n=3.15\pm 0.03$ (Archibald et al. 2016a). It is very challenging to understand the existence of both braking index larger than three and smaller than three. Previously, the braking index of eight pulsar smaller than three can be understood in the wind braking model (Ou et al. 2016). By considering the coupled evolution of rotation and inclination angle, it is shown that both the high and low braking indices of pulsars can be reproduced in the wind braking model. The pulsar braking index is expected to evolve from larger than three to about one.

The wind braking model is based on Kou \& Tong (2015). The inclusion of inclination angle evolution is based on the prescription of Philippov et al. (2014). 

\section{Coupled evolution of rotation and inclination angle in the wind braking model}

\subsection{The wind braking model of pulsars considering the evolution of inclination angle}

The pulsar is generally an oblique rotator. The rotational evolution equation is (Michel \& Goldwire 1970; Philippov et al. 2014):
\begin{equation}
\label{torque}
I\frac{d\boldsymbol{\Omega}}{dt}=\boldsymbol{K},
\end{equation}
where $I=10^{45} \rm \,g \,cm^{2}$ is the moment of inertia, $\boldsymbol{\Omega}$ is the angular velocity, and 
$\boldsymbol{K}$ is the torque working on the star. For a spherical system\footnote{For a non-spherical star, the torque may also lead the rotational axis deviate from the rotational-magnetic plane (i.e., precession) . For the sake of simplicity, a spherical system is assumed.}, equation (\ref{torque}) can be expressed as:
\begin{equation}
\label{z-torque}
I \frac{d\Omega}{dt}=K_{\rm spinning},
\end{equation}
\begin{equation}
\label{x-torque}
I \Omega \frac{d\alpha}{dt}=- K_{\rm alignment},
\end{equation}
where $\alpha$ is the angle between magnetic axis and rotational axis, (i. e., the inclination angle), $K_{\rm spinning}$ and $K_{\rm alignment}$ are the projections of the torque to brake down the pulsar and align the rotational and magnetic axes, respectively.

In the wind braking model, the rotational energy is consumed by the magnetic dipole radiation and particle acceleration. Equation (\ref{z-torque}) can be written as (Xu \& Qiao 2001; Kou \& Tong 2015):
\begin{equation}
\label{rotation}
I \frac{d\Omega}{dt}=-\frac{2\mu^{2} \Omega^{3}}{3 c^{3}} (\sin^2 \alpha +3 \kappa \frac{\Delta{\phi}}{\Delta\Phi})
\equiv -\frac{2\mu^{2} \Omega^{3}}{3 c^{3}} \eta,
\end{equation}
where $\mu=1/2BR^{3}$ is the magnetic dipole moment ($B$ is the magnetic field and $R$ is the neutron star radius), $c$ is the speed of light, $\kappa$ means the primary particle density\footnote{In the wind braking model, all the particles injected into the magnetosphere from the acceleration region are defined as the primary particles. The ``primary particles'' are relative to the ``secondary particles'' which are generated subsequently and responsible for the radio emission.} is $\kappa$ times the Goldreich$-$Julian charge density (Goldreich \& Julian 1969), $\Delta\phi$ is the acceleration potential in the acceleration gap, and $\Delta\Phi=\mu \Omega^{2}/c^{2}$ is the maximum potential for a rotating dipole (Ruderman \& Sutherland 1975). $\eta$ is a dimensionless function. It can be viewed as the dimensionless spin-down torque. The expressions of $\eta$ for different acceleration models are described in the Table 2 of Kou \& Tong (2015). The vacuum gap model (Ruderman \& Sutherland 1975) is taken as an example: $\eta^{\rm CR}_{\rm VG}=\sin^{2} \alpha +4.96 \times 10^{2} \kappa B_{\rm 12}^{-8/7}\Omega^{-15/7}$, where $B_{\rm 12}$ is the magnetic field in units of $10^{12} \rm \, G$. Compared with the previous wind braking model, a $\cos^{2}\alpha$ factor is omitted. Phenomenologically, the $\cos^{2} \alpha$ is a weighting factor between the magnetic-dipole radiation and particle wind. Besides, considering the result of the magnetospheric simulations (Li et al. 2012), the $\cos^{2}\alpha$ factor may not appear in the particle wind component.  

Because the two components of the spin-down torque are independent of the inclination angle when the rotational axis is respectively vertical ($\alpha=90^{\circ}$) and parallel ($\alpha=0^{\circ}$) to the magnetic axis, the spin-down torque and the alignment torque are related as: $ K_{\rm alignment}=[K_{\rm spinning}(0^{\circ})-K_{\rm spinning}(90^{\circ})]\sin\alpha \cos\alpha $ (Philippov et al. 2014). Then equation (\ref{x-torque}) can be written as:
\begin{equation}
\label{alignment}
I \Omega \frac{d\alpha}{dt}=-\frac{2\mu^{2} \Omega^{3}}{3c^{3}}\sin\alpha \cos\alpha.
\end{equation}
The form of alignment torque in the wind braking model is the same as that in the vacuum magnetosphere and similar to that in the magnetohydrodynamical (MHD) simulation (Philippov et al. 2014). 

For the long-term evolution of pulsars, the effect of pulsar death should be considered (Zhang et al. 2000; Contopoulos \& Spitkovsky 2006). The detailed treatment of pulsar death can be found in Kou \& Tong (2015). 

\subsection{Coupled evolution of rotation and inclination angle}

\begin{figure}[!t]
\centering
\begin{minipage}{0.45\textwidth}
\includegraphics[height=0.25\textheight]{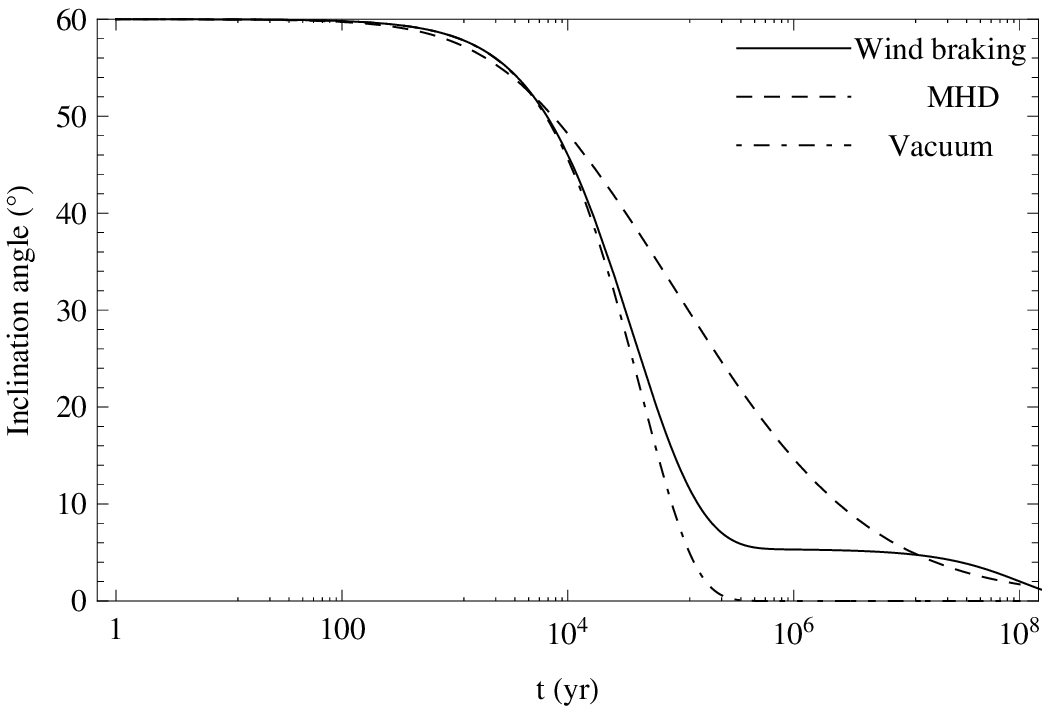}
\end{minipage}
\begin{minipage}{0.45\textwidth}
\includegraphics[height=0.25\textheight]{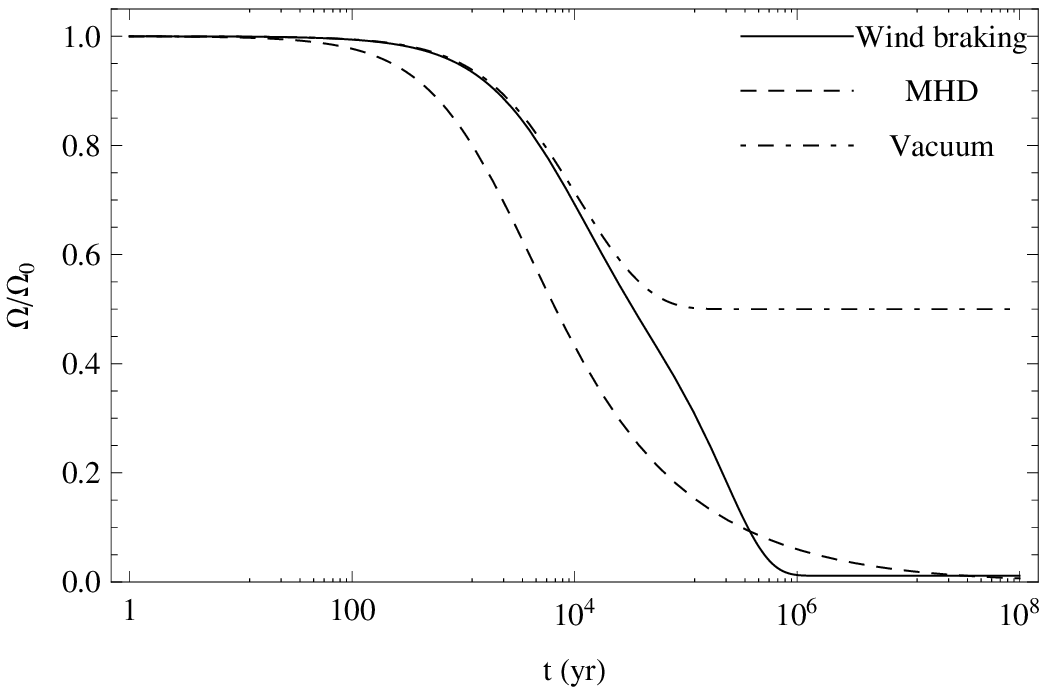}
\end{minipage}
\caption{Evolution of pulsar inclination angle (upper panel) and angular velocity (lower panel) in the wind braking model (solid line). The cases of MHD simulation (dashed line) and a vacuum magnetosphere (dot-dashed line) are also shown for comparison.}
\label{galphaomega}
\end{figure}

To compare with the calculations of vacuum magnetosphere (i.e., magnetic dipole braking) and MHD simulation, the fiducial initial period $P_{0}=10 \rm \, ms$, initial inclination angle $\alpha_{0}=60^{\circ}$ and magnetic field $B=10^{12}\rm \, G$ are assumed. The same parameters are also used for the wind braking model. Besides, $\kappa=100$ are used in the wind braking model. The primary particle density of young pulsars is at least $80$ times of Goldreich-Julian charge density in the vacuum gap model (Kou \& Tong 2015; Ou et al. 2016). A much larger particle density than the Goldreich-Julian density in the pulsar magnetosphere is also found in other models and observations (see discussions in Kou \& Tong 2015 and references therein). Figure \ref{galphaomega} (upper panel) shows the evolution of pulsar inclination angle. The inclination angle tends to align in these models. Compared with the vacuum magnetosphere model, the evolution rate of inclination angle in the plasma-filled magnetosphere (MHD simulation and wind braking model) is smaller, which means that the particle will delay the alignment of pulsar inclination angle. 
Figure \ref{galphaomega} (lower panel) shows the angular velocity evolution. In the vacuum magnetosphere, the spin-down behavior tends to stop and the angular velocity tend to a constant value when the inclination angle is very small. In the MHD simulation and wind braking model, the angular velocity evolution tends to $0$. This is due to the presence of an additional spin-down torque even in the case of very small inclination angle. The difference between the MHD simulation and wind braking model is due to the different form of the spin-down torque (equation (\ref{rotation})  in this paper and equation ($16$) in Philippov et al. 2014). 

\begin{figure}[!t]
\centering
\includegraphics[width=0.45\textwidth]{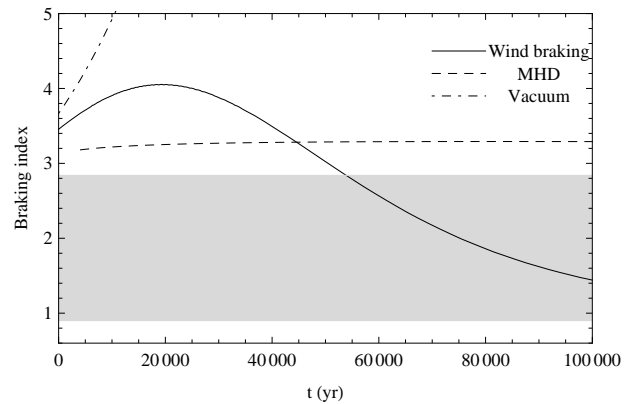}
\caption{Evolution of braking index in the wind braking model (solid line). The case of MHD simulation (dashed line) and a vacuum magnetosphere (dot-dashed line) are also shown for comparison. The shaded region represents the range of observed braking indices of young pulsars ($0.9-2.84$, Lyne et al. 2015).}
\label{gnt}
\end{figure}

The evolution of pulsar braking index is shown in Figure \ref{gnt}. The braking index expected in the vacuum magnetosphere and MHD simulation is exactly $3$ when not considering the evolution of pulsar inclination angle (Philippov et al. 2014). And because of the inclination angle alignment, braking index in the vacuum magnetosphere will always be larger than $3$ and the line evolves quickly. Due to the effect of particles in the magnetosphere, braking index evolution in the MHD simulation is much gentle. However, braking index in this case is also always larger than $3$ (Arzamasskiy et al. 2015). Braking index observations of smaller than $3$ can be explained in the wind braking model. Furthermore, the inclination angle alignment will also affect the braking index evolution in the wind braking model. According to the definition of pulsar braking index, the expression of pulsar braking index when considering the evolution of inclination angle is:
\begin{equation}
\label{brakingindex}
n=3+\frac{\Omega}{\eta}\frac{d\eta}{d\Omega}+\frac{\tau_{c}}{\tau_{\alpha}} \frac{\alpha}{\eta} \frac{d \eta}{d \alpha},
\end{equation}
where $\tau_{c}=-\frac{\Omega}{2\dot{\Omega}}$ is the characteristic age and $\tau_{\alpha}=-\frac{\alpha}{2\dot{\alpha}}$ is the evolution timescale of inclination angle. An aligning inclination angle ($\tau_{\alpha}>0$) will lead to a larger braking index. The braking index will increase even larger than $3$ during the early time when magneto-dipole radiation dominates the spin-down torque. However, as the wind component begins to dominate the pulsar spin down behavior, the pulsar braking index will decrease. And during the following long time, the braking index will be smaller than $3$. The minimum braking index is $6/7$ in the vacuum gap model (Ou et al. 2016). Generally, the minimum braking index is about one in the wind braking model. 

\begin{figure}[!t]
\centering
\includegraphics[width=0.45\textwidth]{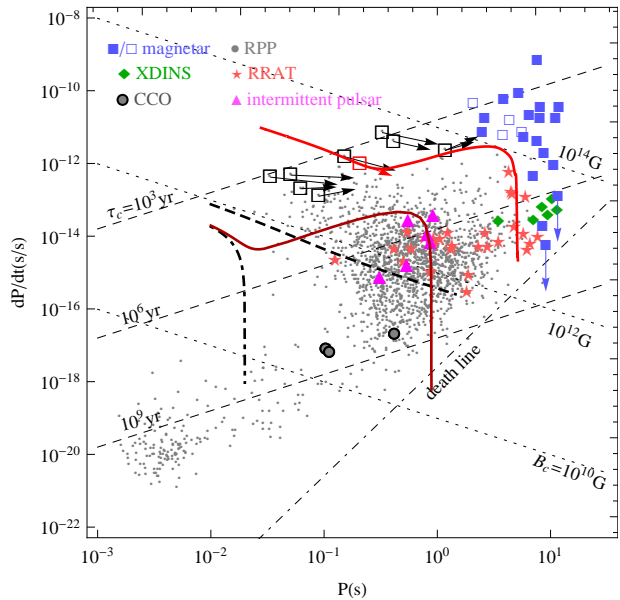}
\caption{Long-term evolution of pulsars in the $P-\dot{P}$ diagram. The dark red line is for the wind braking model. The MHD case (dashed line) and a vacuum magnetosphere (dashed line) are also shown for comparison. The red empty square is the observations of PSR J1640$-$4631. And the arrow represents its evolution direction in the next $10^4$ years. The red line is the evolution track of PSR J1640$-$4631, assuming a present inclination angle of $45^{\circ}$. The black squares and their arrows mark the evolution direction of the eight young pulsars with braking index measured. The $P-\dot{P}$ diagram of pulsars is updated from Fig. 6 in Kou \& Tong (2015).}
\label{gPPdot}
\end{figure}

Figure \ref{gPPdot} shows the long-term evolution of pulsars in the $P-\dot{P}$ diagram. In the vacuum magnetosphere, the evolution line drops quickly before the group of rotation powered pulsars. The line of the MHD simulation evolves towards down-right and through the rotation powered pulsar group. In the wind braking model, the pulsar evolves firstly down-right in the magneto-dipole radiation dominated case, then up-right in the particle wind dominated case and eventually go down due to the effect of pulsar ``death''. Fron the definition of pulsar braking index, the period-derivative and period are related as: $\dot{P} \propto P^{2-n}$ (Espinoza et al. 2011).  The evolution from down-right to up-right of pulsars in the $P-\dot{P}$ diagram indicates the evolution of pulsar braking index from 3 to about 1. This is consistent with the results of figure \ref{gnt}.

\subsection{Calculations for the high braking index pulsar PSR J1640$-$4631}

A high braking index ($>3$) is claimed for PSR J1640$-$4631 (Archibald et al. 2016a)\footnote{If this braking index is rejected by future observations, the main conclusion of this paper will remain unchanged, except for the calculations in this subsection.}. Given the timing observations $\nu=4.843 \rm\, s^{-1}$, $\dot{\nu}=-2.28 \times 10^{-11} \rm\, s^{-2}$, $n=3.15 (3)$, and an assumed present inclination angle\footnote {There is no observational or best fitted inclination angle given. Three typical inclination angles of $15^{\circ}$, $45^{\circ}$ and $70^{\circ}$ are choosen.} of PSR J1640$-$4631, the  magnetic field, particle density and the evolution rate of inclination angle can be calculated by equations (\ref{rotation}), (\ref{alignment}) and (\ref{brakingindex}). And assuming a present age for PSR J1640$-$4631 ($3000 \rm \, yr$ for present inclination angle of $45^{\circ}$ and $70^{\circ}$ and $2000\rm \, yr$ for $15^{\circ}$), the initial spin period and inclination angle can be calculated by integrating equation (\ref{rotation}) and (\ref{alignment}). Table \ref{parameters of J1640} shows the parameters of PSR J1640$-$4631 in the wind braking model. Comparably, the values of $\kappa$ are smaller. This means that the proportion of particle wind of PSR J1640$-$4631 is relatively weak. The effect of inclination angle alignment on braking index can not be covered by the particle wind effect. Hence, its braking index will be lager than $3$ at present. The present alignment rate of the inclination angle is $-0.56 \rm \,^{\circ}/century$ if $\alpha_{\rm present}=45^{\circ}$. 

In the wind braking model, the predicted frequency third derivative is $\stackrel{...}{\nu} \approx -10^{-32} \,\rm s^{-4}$, corresponding to a second braking index of $m \equiv \nu^2 \stackrel{...}{\nu} /\dot{\nu}^3 \approx 18$. It is consistent with the present upper limits of $|\stackrel{...}{\nu}| < 1.4\times 10^{-30} \,\rm s^{-4}$ (Archibald et al. 2016a). 

\begin{figure}[!t]
\centering
\begin{minipage}{0.45\textwidth}
\includegraphics[height=0.25\textheight]{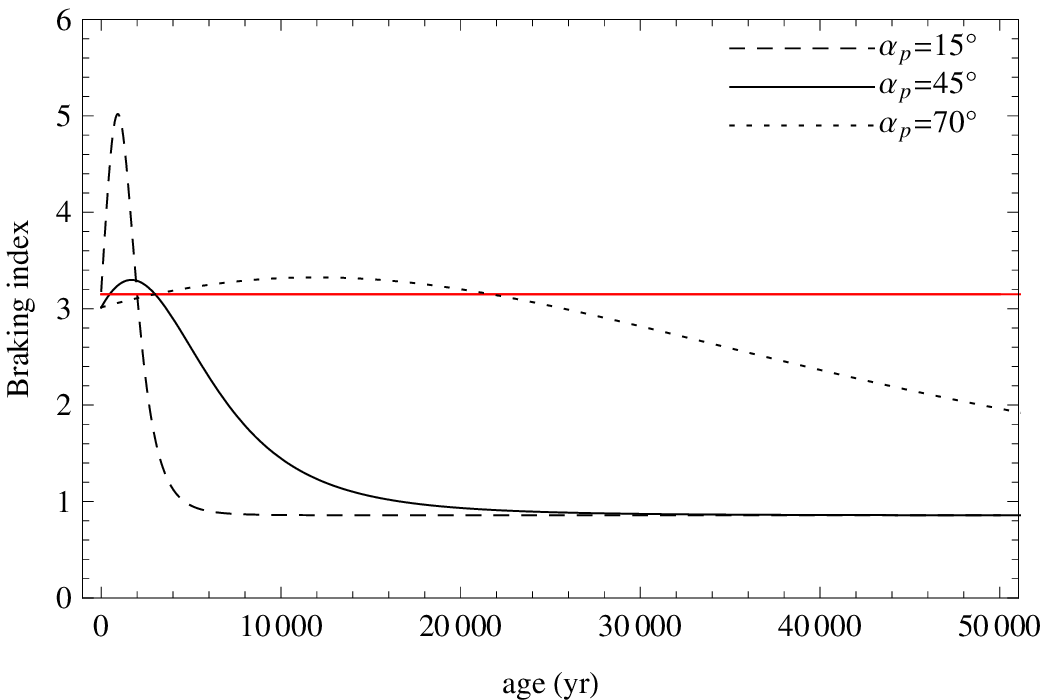}
\end{minipage}
\begin{minipage}{0.45\textwidth}
\includegraphics[height=0.25\textheight]{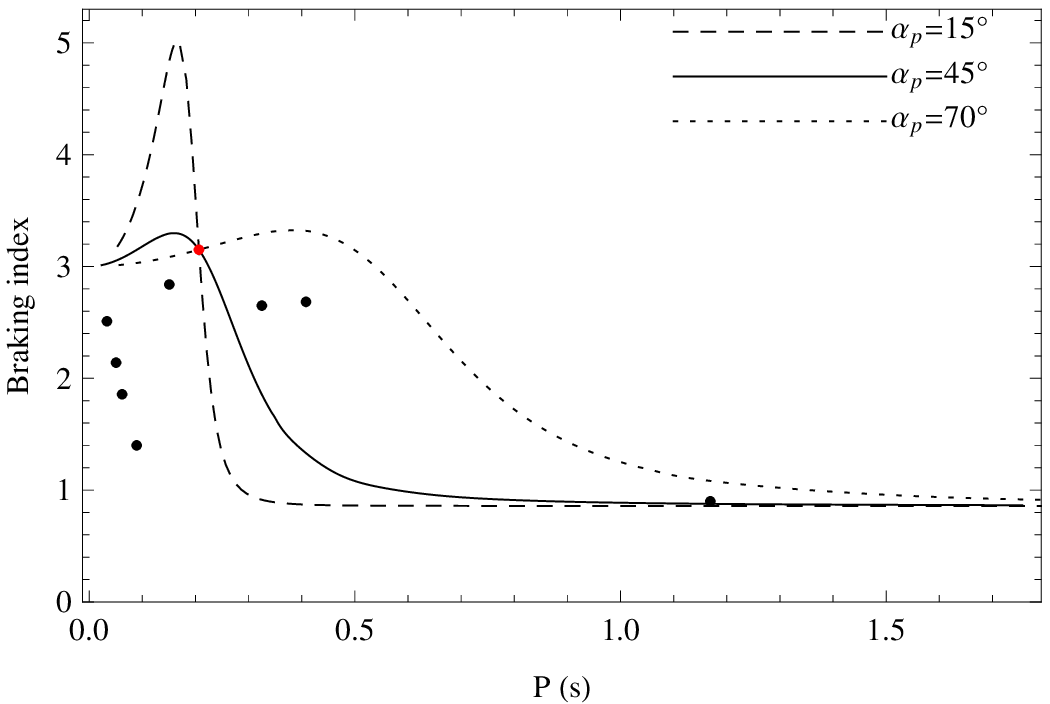}
\end{minipage}
\caption{The braking index evolution of PSR J1640$-$4631 as a function of time (upper panel) and as a function of rotational period (lower panel). Different present inclination angles of $15^{\circ}$, $45^{\circ}$ and $70^{\circ}$ are shown. The red solid line (upper panel) and the red point (lower panel) are the observations of PSR J1640$-$4631 (Archibald et al. 2016a). The black points (lower panel) are the braking index observations of eight young pulsars (Lyne et al. 2015). }
\label{gnt1640}
\end{figure}

Figure \ref{gnt1640} (upper panel) shows the braking index evolution of PSR J1640$-$4631 with time. In the early time, the braking index will increase even lager than $3$ because of the inclination angle alignment. As the particle wind begins to dominate the spin-down behavior, its braking index will decrease. And during the following long time, its braking index will be smaller than $3$. Figure \ref{gnt1640} (lower panel) shows the braking index evolution  of PSR J1640$-$4631 as a function of rotational period. Compared with other young pulsars, PSR J1640$-$4631 lies in the magneto-dipole radiation dominated case and its present braking index will be larger than $3$. But as the pulsar spin-down, the effect of particle wind will be more and more important, its braking index will decrease. The red line in figure \ref{gPPdot} shows the evolution of PSR J1640$-$4631 in the $P-\dot{P}$ diagram.

\begin{table}[!t]
\begin{center}
\caption{Parameters of PSR J1640$-$4631 in the wind braking model.}
\label{parameters of J1640}
\begin{tabular}{llllll}
\hline\hline
$\alpha$  & $B_{12}$ & $\kappa$ & $\dot{\alpha}$  & $P_{0}$  & $\alpha_{0}$ \\
$(^{\circ})$ & $(10^{12}\rm \, G)$ &  & $(^{\circ}/\rm century)$ & $(\rm \, ms)$ & $(^{\circ})$ \\
\hline
$15$ & $55$ & $60$ & $-0.8$ & $53$ & $72$\\
$45$ & $33$ & $42$ & $-0.56$ & $23$ & $85$\\
$70$ & $30$ & $6$ & $-0.3$ & $57$ & $84$\\

\hline

\end{tabular}
\end{center}
Notes: $\alpha$ is the assumed present inclination angle; $B_{12}$ is the magnetic field in unit of $10^{12}\rm\,G$; $P_{0}$ and $\alpha_{0}$ are the initial rotational period and inclination angle, respectively. 
\end{table}

\subsection{Comparison with the MHD simulations}

The wind braking model considers the pulsar as an oblique rotator. The magnetic dipole moment has both a perpendicular component and  a parallel component, perpendicular and parallel to the rotational axis. The parallel component may be responsible for the particle acceleration (Ruderman \& Sutherland 1975). The perpendicular component may be approximated by the magnetic dipole radiation. Therefore, as an educated guess, the pulsar spin-down torque (equation \ref{rotation})  is made up of two components (Xu \& Qiao 2001). The key input is the particle component, which provides a natural link between the timing and emission properties of pulsars. In the presence of a particle component, the pulsar braking index lies between three and one. 

Considering recent progresses of MHD simulations of pulsar magnetospheres (Spitkovsky 2006; Li et al. 2012; Kalapotharakos et al. 2012; Contopoulos et al. 2014; Philippov et al. 2014), a general form of pulsar spin-down torque is 
\begin{eqnarray}
\nonumber
I \frac{d\Omega}{dt}&=& - k_1 \frac{\mu^{2} \Omega^{3}}{ c^{3}} (\sin^2 \alpha +\text{``particle term''}) \\
&\equiv&- k_1 \frac{\mu^{2} \Omega^{3}}{ c^{3}} \eta,
\end{eqnarray}
where $k_1$ is a numerical factor.  Mathematically, the term proportional to $\sin^2\alpha$ may be dubbed as the ``dipole term''. While, the remaining term may be dubbed as the ``particle term''. In the wind braking model, the numerical factor is $k_1 =2/3$. In the MHD simulations, the numerical factor is $k_1 \approx 1$
(Spitkovsky 2006; Philippov et al. 2014).  However, according to the ``new standard pulsar magnetosphere'' (Contopoulos et al. 2014), the numerical factor is $k_1 \approx 0.82$. Particle-in-cell simulations also found a numerical factor smaller than one (Philippov et al. 2015), because particle acceleration means the presences of vacuum regions in the pulsar magnetosphere. Furthermore, the difference between the numerical factor of the wind braking model and the MHD simulations will mainly affect the magnetic field strength. Therefore, a different numerical factor will not affect the conclusions here. 

In the wind braking model, the ``particle term'' is $3 \kappa \Delta\phi/\Delta\Phi$. It is determined by the number of outflow particles and the acceleration potential. If the acceleration potential is assumed to be equal to the maximum acceleration potential, and the particle number density is assumed to be equal to the Goldreich-Juilan density, then the ``particle term'' in the wind braking model is $3$. Considering a numerical factor of $2/3$, then the particle term is $2$. In the MHD simulation, the ``particle term'' is $1$. Therefore, the ``particle term'' in the MHD simulations can be deduced from the wind braking model, by assuming a maximum acceleration potential and a Goldreich-Julian particle density. The only difference is a factor of two. In the resistive MHD simulations (Li et al. 2012; Kalapotharakos et al. 2012), the particle term is slightly modified, see equation (13) in Li et al. (2012). It is similar to that in Contopoulos \& Spitkovsky (2006): $\sin^2\alpha+ (1-V_{\rm drop}/V_{\rm pc})$, except for a different combination of angular factor. If the acceleration potential is assumed to be equal to the maximum acceleration potential, the corresponding braking index is always equal to 3. Considering the effect of pulsar death or inclination angle evolution, the braking index will be larger than 3. However, in physical acceleration models for pulsar magnetospheres, the acceleration potential is different from the maximum acceleration potential (Xu \& Qiao 2001 and references therein). Recent observations and modeling also show possible evidence of a much higher particle density in the pulsar magnetosphere (Kou \& Tong 2015 and references therein).  By considering these two aspects, the corresponding expression of wind braking model is obtained. 

In the original version of the wind braking model (Xu \& Qiao 2001), the dimensionless torque is $\eta = \sin^2\alpha+  ... \times \kappa\cos^2\alpha$. An additional angular factor $\cos^2\alpha$ is present. This kind of combination is also found in Contopoulos \& Spitkovsky (2006). In previous works of wind braking of pulsars, the possible evolution of pulsar inclination angle is not considered and the inclination angle is constant. Since the particle density and inclination angle always appeared as $\kappa \cos^2\alpha$, for a constant inclination angle, there is degeneracy between the particle number density $\kappa$ and $\cos^2\alpha$ (Rogers \& Safi-Harb 2017). If there is no $\cos^2\alpha$ in the $\eta$ function, all the previous results can be obtained by replacing the corresponding particle density $\kappa$ by $\kappa \cos^2\alpha$. This kind of ambiguity is already known when applying the wind braking model to intermittent pulsars (Li et al. 2014). In order to keep consistent with previous works, the $\cos^2\alpha$ factor is kept. When not considering the evolution of pulsar inclination angle, this ambiguity will not affect the physical results. According to the relation between pulsar spin-down and alignment (Philippov et al. 2014), the inclination angle evolves to decrease the spin-down torque. If there is a $\cos^2\alpha$ in the dimensionless torque, then: (1) in the dipole radiation dominated case ($\eta$ is dominated by the dipole term which is proportional to $\sin^2\alpha$), the inclination will decrease will time. (2) As the pulsar evolves, the particle component begins to dominate ($\eta$ is dominated by the particle term which is proportional to $\cos^2\alpha$) and the inclination angle will increase with time. However, statistically the pulsar inclination angle tends to decrease with time (Lyne \& Manchester 1988; Tauris \& Manchester 1998). Therefore, compared with the observations of pulsar inclination angle, the possible $\cos^2\alpha$ factor may not appear.

The evolution of inclination angle in the wind braking model (equation \ref{alignment}) is done in analogy with that of magnetic dipole braking and MHD simulations. Apart from a different numerical factor, the alignment equation in the magnetic dipole case and MHD case can be viewed as the same (equation (7) and equation (15) in Philippov et al. 2014). This is because they have the same dipole term, the term proportional to $\sin^2\alpha$. The term independent of the inclination angle in the spin-down torque does not contribute to the alignment torque. Since the wind braking model has the same dipole term, it is possible that the equation for inclination angle evolution is also the same. Furthermore, according to the prescription of Philippov et al. (2014, equation (33) there), equation (\ref{alignment}) can be deduced when assuming that the coefficient A and B are independent of the inclination angle. Even if A or B depends on the inclination angle, the inclination angle will always decrease with time for the spin-down torque of equation (\ref{rotation}). 
The resulting change is only quantitative. 
At the early age, when the magnetic dipole radiation dominates the spin-down torque, for a decreasing inclination angle, the corresponding braking index may also be larger than three. 

It is tempting to combine the MHD simulations with some amount of particle outflow. Then the dimensionless spin-down torque will be of the form: $\eta = 1+ \sin^2\alpha + \text{``particle term''}$. For young pulsars with braking index measured, their inclination angle may not have decreased significantly, see figure \ref{galphaomega}. For an inclination angle of $45^{\circ}$, in the particle wind dominated case, the ``particle term'' will be significantly larger than the dipole term which is proportional to $\sin^2\alpha =1/2$. Then if the dipole term is replaced by the MHD results, the ``particle term'' will also be larger than the MHD term which is proportional to $1+\sin^2\alpha=3/2$. Therefore, mathematically, such a combination will not affect the braking index evolution of young pulsars\footnote{For old pulsars, such a combination will provide an additional spin-down torque when the inclination angle becomes very small.}. Numerical calculations also confirm this analysis. Furthermore, when the particle acceleration is introduced in some way in magnetospheric simulations (Li et al. 2012; Kalapotharakos et al. 2012; Philippov et al. 2015), such a combination of ``MHD + particle term'' is not found. The main modification is the term ``1'' in the MHD simulations (Li et al. 2012). This may tells us that this term ``1'' is associated with particle accelerations. 
Therefore, the results of resistive magnetospheric simulations are not in strong support for such a combination. However, such a combination cannot be ruled out at present. The study of pulsar braking index is not very sensitive to discriminate between the two cases of ``dipole+particle wind'' or ``MHD+particle wind''. Both a ``$\sin^2\alpha$'' and ``$1+\sin^2\alpha$'' term will result in a braking index of three. Other pulsar observations may help to solve this problem, e.g. intermittent pulsars etc. 

In summary, the importance of a particle outflow is stressed in the wind braking model . This particle term will result in a braking index smaller than three. Compared with the results of MHD simulations, there may be many assumptions in the wind braking model. However, these uncertainties will not affect the final conclusions. If future pulsar magnetosphere simulations can model the particle acceleration and injection into the magnetosphere more physically, the basic assumptions of the wind braking model can be tested.

\section{Discussion and conclusion}

Lyne \& Manchester (1988) had analyzed the polarization information of hundreds of pulsars. The statistical studies showed that the inclination angle distribution for young pulsar is uniform but aligned for older pulsars (Lyne \& Manchester 1988). By studying two groups of pulsar polarization data, Tauris \& Manchester (1998) made a further studies and they concluded that the pulsar inclination angle tends to align. The inclination angle evolution in the wind braking model is consistent with these observations. Lyne et al. (2013) proposed that the inclination angle of the Crab pulsar is increasing with an increasing rate $0.62^{\circ}\pm0.03^{\circ}$ per century. On the one hand, an increasing inclination angle is one of the possible explanations to the steady increase in the separation of the main pulse and interpulse  of the Crab pulsar (Lyne et al. 2013). On the other hand, the increasing inclination angle may be caused by the pulsar precession in the non-spherical case (Arzamasskiy et al. 2015).

There are also other works to explain the observed high braking index of PSR J1640$-$4631. Braking index will be larger than $3$ in the plasma filled magnetosphere because of the inclination angle alignment (Eksi et al. 2016; and figure \ref{gnt} in this paper) . And the possible gravitational wave emission will also lead to a high pulsar braking index (de Araujo et al. 2016; Chen et al. 2016). However, these works only try to explain the braking index observations of larger than $3$. 

In the wind braking model, the braking index in the early time can be larger than three\footnote{The braking index can be as high as $5$, depending on the parameter space, see figure \ref{gnt1640}.} when considering the possible evolution of inclination angle. At later time, it will evolve from about three to about one. Therefore, a general anticipation is that pulsars with higher braking index should be younger than those with lower braking index. The pulsar braking index should evolve from PSR J1640$-$4631-like case  (larger than three), to that of Crab-like case (about three), then to Vela-like case (about one). There is marginal evidence for the braking index evolution (Espinoza  2013). At later time, the effect of pulsar death will lead the pulsar to the death valley (figure \ref{gPPdot}). A very large braking index may be expected in this case. However, for these old pulsars, the fluctuation in magnetosphere will dominate the frequency second derivatives. And the observed $\ddot{\nu}$ may be dominated by the fluctuation of the magnetosphere (i.e., timing noise, Ou et al. 2016). 

The possible evolution of pulsar braking index discussed above are for the long term evolutions. Two kinds of short term evolutions of pulsar braking index are possible (Kou et al. 2016): (1) A discrete change of particle density. For a higher particle density, the pulsar is expected to have a higher spin-down rate, and lower braking index. (2) A secular change of particle density. An increasing particle density will result in a lower braking index, while not affecting the spin-down rate significantly. Case (2) may correspond to the lower braking index of PSR J1846$-$0258 (Archibald et al. 2015). The state change and smaller braking index of PSR B0540$-$69 (Marshall et al. 2015,2016) are consistent with Case (1). The much smaller braking index of PSR B0540$-$69 than predicted may be due to secular changes in the particle density, similar to that of PSR J1846$-$0258. A general prediction in the wind braking model is that: when the magnetospheric activities are stronger, the braking index will be smaller. 
A smaller braking index of PSR J1119$-$6127 is expected after its outburst (Archibald et al. 2016b). 

In conclusion, both braking index higher than three and the small braking index of pulsars can be obtained in the wind braking model, by including the evolution of inclination angle. No additional braking mechanism is required. The pulsar braking index is expected to evolve from larger than three to about one. Future observations of more sources will help to make clear the possible long term and short term evolutions of pulsar braking index. 

Notes added: During the reviewing process, more pulsars with braking index measurement are reported. Gamma-ray timing of a young pulsar found a braking index of $n=2.598$ (Clark et al. 2016). Vela-like glitching pulsars generally have braking indices of $n\le 2$ (Espinoza et al. 2017). These observations are consistent with the prediction in the wind braking model, i.e. braking index evolves from the Crab-like case to the Vela-like case.

\section*{Acknowledgments}
The authors would like to thank H. G. Wang for discussions. 
H.Tong is supported by 973 Program (2015CB857100), and NSFC (U1531137).


\begin{thebibliography}{99}

\bibitem{Archibald2015}
Archibald, R. F., Kaspi, V. M., Beardmore, A. P., et al. 2015, ApJ, 810, 67

\bibitem{Archibald2016a}
Archibald, R. F., Getthelf, E. V., Ferdman, R. D., et al. 2016a, ApJL, 819, L16

\bibitem{Archiblda2016b}
Archibald, R. F., Kaspi, V. M., Tendulkar, S. P., \& Scholz, P. 2016b, ApJL, 829, L21

\bibitem{Arzamasskiy2015}
Arzamasskiy, L., Philippov, A., \& Tchekhovskoy, A. 2015, MNRAS, 453, 3540

\bibitem{Chen2016}
Chen, W. C. 2016, A\&A, 593, L3

\bibitem{Clark2016}
Clark, C. J., Pletsch, H. J., Wu, J., et al. 2016, ApJL, 832, L15

\bibitem{Contopoulos2006}
Contopoulos, I., \& Spitkovsky, A. 2006, ApJ, 643, 1139

\bibitem{Contopoulos2014}
Contopoulos, I., Kalapotharakos, C., \& Kazanas, D. 2014, ApJ, 781, 46

\bibitem{deAraujo2016}
de Araujo, J. C. N., Coelho,  J. G., \& Costa, C. A. 2016, JCAP, 7, 023

\bibitem{Eksi2016}
Eksi, K. Y., Andac, I. C., Cikintoglu, S., et al. 2016, ApJ, 823, 34

\bibitem{Espinoza2011}
Espinoza, C. M., Lyne, A. G., Kramer, M., et al. 2011, ApJL, 741, L13

\bibitem{Espinoza2013}
Espinoza, C. M. 2013, IAUS, 291, 195 (arXiv:1211.5276)

\bibitem{Espinoza2017}
Espinoza, C. M., Lyne, A. G., \& Stappers, B. W. 2017, MNRAS, 147, 162

\bibitem{Goldreich1969}
Goldreich, P., \& Julian, W. H. 1969, ApJ, 157, 869

\bibitem{Kalapotharakos2012}
Kalapotharakos, C., Kazanas, D., Harding, A., \& Contopoulos, I. 2012, ApJ, 749, 2

\bibitem{Kou2015}
Kou, F. F., \& Tong, H. 2015, MNRAS, 450, 1990

\bibitem{Kou2016}
Kou, F. F., Ou, Z. W., \& Tong, H. 2016, RAA, 16, 10

\bibitem{Li2012}
Li, J., Spitkovsky, A., \& Tchekhovskoy, A. 2012, ApJ, 746, 60

\bibitem{Li2014}
Li, L., Tong, H., Yan, W. M., et al. 2014, ApJ, 788, 16

\bibitem{Liu2014}
Liu, X. W., Xu, R. X., Qiao, G. J., et al. 2014, RAA, 14, 85

\bibitem{Lyne1988}
Lyne, A. G., \& Manchester, R. N. 1988, MNRAS, 234, 477

\bibitem{Lyne2013}
Lyne, A., Graham-Smith, F., Weltevrede, P., et al. 2013, Science, 342, 598

\bibitem{Lyne2015}
Lyne, A. G., Jordan, C. A., Graham-Smith, F., et al. 2015, MNRAS, 446, 857

\bibitem{Marshall2015}
Marshall, F. E., Guillemot, L., Harding. A. K., et al. 2015, ApJL, 807, L27

\bibitem{Marshal2016}
Marshall, F. E., Guillemot, L., Harding. A. K., et al. 2016, ApJL, 827, L39

\bibitem{Michel1970}
Michel, F. C., \& Goldwire, Jr. H. C. 1970, ApL, 5, 21

\bibitem{Ou2016}
Ou, Z. W., Tong, H., Kou, F. F., \& Ding G. Q. 2016, MNRAS, 457, 3922

\bibitem{Philippov2014}
Philippov, A., Tchekhovskoy, A., \& Li, J. G. 2014, MNRAS, 441, 1879

\bibitem{Philippov2015}
Philippov, A. A., Spitkovsky, A., \& Cerutti, B. 2015, ApJL, 801, L19

\bibitem{Rogers2017}
Rogers, A., \& Safi-Harb, S. 2017, MNRAS, 465, 383

\bibitem{Ruderman1979}
Ruderman, M. A., \& Sutherland, P. G. 1975, ApJ, 196, 51

\bibitem{Spitkovsky2006}
Spitkovsky, A. 2006, ApJL, , 648, L51

\bibitem{Tauris1998}
Tauris, T. M., \& Manchester, R. N. 1998, MNRAS, 298, 625

\bibitem{Xu2001}
Xu, R. X., \& Qiao, G. J. 2001, ApJL, 561, L85

\bibitem{Zhang2000}
Zhang, B., Harding, A. K., \& Muslimov, A. G. 2000, ApJL, 531, L135

\end{thebibliography}
\end{document}